\documentclass[ALICE,manyauthors]{cernphprep}
\usepackage[usenames, dvipsnames]{color}
\usepackage[comma,square,numbers,sort&compress]{natbib}
\usepackage{hyperref}
\usepackage{lineno}
\def\snn{$\mathbf{\sqrt{\textit{s}_{_{\rm NN}}}}$}
\def \pb{{Pb--Pb}}

\newcommand{ \be }{\begin{eqnarray}}
\newcommand{ \ee }{\end{eqnarray}}

\newcommand{ \bk }{{\bf k}}
\newcommand{ \bp }{{\bf p}}

\newcommand{ \bq }{{\bf q}}

\newcommand{ \rT }{{\rm T}}

\newcommand{\pt}{$p_{\rm{T}}$~}
\newcommand{\kt}{k_{\rm{T}}}
\newcommand{\gevc}{{\rm GeV/{\it c}}}
\newcommand{\rside}{R_{\rm side}}
\newcommand{\rout}{R_{\rm out}}
\newcommand{\rlong}{R_{\rm long}}
\newcommand{\ros}{R_{\rm os}}
\newcommand{\rol}{R_{\rm ol}}
\newcommand{\rsl}{R_{\rm sl}}
\newcommand{\qside}{q_{\rm side}}
\newcommand{\qout}{q_{\rm out}}
\newcommand{\qlong}{q_{\rm long}}

 \newcommand{\dphi}{\Delta\varphi}
\newcommand{\PsiTh}{\Psi_{\mathrm{EP},3}}

\makeatletter
\def\NAT@def@citea{\def\@citea{\NAT@separator}}
\makeatother

\begin{document}%

%%%%%%%%%%%%%%%  Title page %%%%%%%%%%%%%%%%%%%%%%%%
\begin{titlepage}
\PHyear{2018}
\PHnumber{035}      % required, will be obtained from PH
\PHdate{6 March}  % required, will be obtained from PH
%

%%% Put your own title + short title here:
\title{Azimuthally-differential pion femtoscopy relative to
  the\\ third harmonic event plane in \pb~collisions at
  $\mathbf{\sqrt{\textit{s}_{_{\rm NN}}}}$~= 2.76~TeV }

\ShortTitle{Azimuthally-differential pion femtoscopy in
  \pb~collisions}   % appears on right page headers

%%% Do not change the next lines
\Collaboration{ALICE Collaboration\thanks{See
    Appendix~\ref{app:collab} for the list of collaboration
    members}} \ShortAuthor{ALICE
  Collaboration}
% appears on left page headers, do not change

\begin{abstract}
  Azimuthally-differential femtoscopic measurements, being sensitive
  to spatio-temporal characteristics of the source as well as to the
  collective velocity fields at freeze out, provide very important
  information on the nature and dynamics of the system
  evolution. While the HBT radii oscillations relative to the second
  harmonic event plane measured recently reflect mostly the spatial
  geometry of the source, model studies have shown that the HBT radii
  oscillations relative to the third harmonic event plane are
  predominantly defined by the velocity fields. In this Letter, we
  present the first results on azimuthally-differential pion
  femtoscopy relative to the third harmonic event plane as a function
  of the pion pair transverse momentum $\kt$ for different collision
  centralities in Pb--Pb collisions at \snn~=~2.76 TeV. We find that
  the $\rside$ and $\rout$ radii, which characterize the pion source
  size in the directions perpendicular and parallel to the pion
  transverse momentum, oscillate in phase relative to the third
  harmonic event plane, similar to the results from 3+1D
  hydrodynamical calculations. The observed radii oscillations
  unambiguously signal a collective expansion and anisotropy in the
  velocity fields. A comparison of the measured radii oscillations
  with the Blast-Wave model calculations indicate that the initial
  state triangularity is washed-out at freeze out.
\end{abstract}
\end{titlepage}
\setcounter{page}{2}

\maketitle

%\linenumbers

%===========================================================
\section{Introduction}
\label{sec:intro}
\label{Intro}
Heavy-ion collisions at LHC energies create a hot and dense medium
known as the quark--gluon plasma (QGP)~\cite{Muller:2012zq}. The QGP
fireball first expands, cools, and then freezes out into a collection
of final-state hadrons. Correlations among the particles carry
information about the space--time extent of the emitting source, and
are imprinted on the final-state spectra due to a quantum-mechanical
interference effect~\cite{Goldhaber:1960sf}.  Commonly known as
intensity or Hanbury-Brown--Twiss (HBT) interferometry, the
correlation of two identical particles at small relative momentum, is
an effective tool to study the space--time (``femtoscopic'') structure
of the emitting source in relativistic heavy-ion
collisions~\cite{Bertsch:1988db}.  The initial state of a heavy-ion
collision is characterized by spatial anisotropies that lead to
anisotropies in pressure gradients, and consequently to azimuthal anisotropies
in final particle distributions, commonly called anisotropic
flow. Anisotropic flow is usually characterized by a Fourier
decomposition of the particle azimuthal distribution and quantified by
the flow coefficients $v_{n}$ and the corresponding symmetry plane
angles $\Psi_{n}$~\cite{Poskanzer:1998yz}. Elliptic flow is quantified
by the second flow harmonic coefficient $v_2$, whereas triangular
flow~\cite{Alver:2010gr} is quantified by $v_3$.  Due to the
position--momentum correlations in particle
emission~\cite{Bowler:1986ta}, the particles emitted at a particular
angle relative to the flow plane carry information about the source as
seen from that corresponding direction; these correlations also lead
to the HBT radii to be sensitive to the collective velocity fields,
from which information about the dynamics of the system evolution can
be extracted.

Azimuthally-differential femtoscopic measurements can be performed
relative to the direction of different harmonics event
planes~\cite{Voloshin:1995mc,Voloshin:1996ch}.  The measurements of
the HBT radii with respect to the first harmonic event plane (directed
flow) at the AGS~\cite{Lisa:2000xj} revealed that the source was
tilted relative to the beam direction~\cite{Lisa:2000ip}. The HBT
radii variations relative to the second harmonic event plane angle
($\Psi_2$) provide information on the pion source elliptic
eccentricity at freeze-out.  The recent ALICE
measurements~\cite{Adamova:2017opl} indicate that due to the strong
in-plane expansion the final-state source elliptic eccentricity is
more than a factor 2--3 smaller compared to the initial-state. While
the HBT radii modulations relative to $\Psi_2$ are defined mostly by
the source geometry, the azimuthal dependence of the HBT radii
relative to the third harmonic event plane ($\Psi_3$) originate
predominantly in the anisotropies of the collective velocity fields --
for a triangular, but static source the radii do not exhibit any
oscillations~\cite{Voloshin:2011mg}. Models
studies~\cite{Plumberg:2013nga,Bozek:2014hwa} show that the anisotropy
in expansion velocity as well as the system geometrical shape can be
strongly constrained by azimuthally differential femtoscopic
measurements relative to $\Psi_3$. The HBT radii oscillations relative
to the third harmonic event plane have been first observed in Au--Au
collisions at RHIC energy by the PHENIX
Collaboration~\cite{Adare:2014vax}. Unfortunately, due to large
uncertainties these measurements did not allow to obtain detailed
information on the origin of the observed oscillations.

In this Letter, the first azimuthally-differential femtoscopic
measurement relative to the third harmonic event plane in Pb--Pb
collisions at \snn~=~2.76 TeV from the ALICE experiment are
presented. We compare our results to the toy-model calculations
from~\cite{Plumberg:2013nga} to get an insight on the role of the
anisotropies in the velocity fields and the system shape.  In
addition, we compare our results to a 3+1D hydrodynamical
calculations~\cite{Bozek:2014hwa} and a Blast-Wave Model
\cite{Cimerman:2017lmm} for a quantitative characterization of the final
source shape.

\section{Data analysis}

The analysis was performed over the data sample recorded in 2011
during the second~\pb\ running period at the LHC. Approximately 2
million minimum bias events, 29.2 million central trigger events, and
34.1 million semi-central trigger events were used. The minimum bias,
semi-central, and central triggers used all require a signal in both
V0 detectors~\cite{Abbas:2013taa}.  The V0 detector, also used for the
centrality determination~\cite{Aamodt:2010cz}, is a small angle
detector of scintillator arrays covering pseudorapidity ranges
$2.8<\eta<5.1$ and $-3.7<\eta<-1.7$ for a collision vertex occurring
at the center of the ALICE detector.  The results of this analysis are
reported for collision centrality classes expressed as ranges of the
fraction of the inelastic \pb~cross section: 0--5\%, 5--10\%,
10--20\%, 20--30\%, 30--40\%, and 40--50\%.  Events with the primary
event vertex along the beam direction $|V_{z}|<8$~cm were used in this
analysis to ensure a uniform pseudorapidity acceptance. A detailed
description of the ALICE detector can be found
in~\cite{ALICE_exp_1,Abelev:2014ffa}.  The Time Projection Chamber
(TPC) has full azimuthal coverage and allows charged-particle track
reconstruction in the pseudorapidity range $|\eta|<~0.8$, as well as
particle identification via the specific ionization energy loss
$\mathrm{d}E/\mathrm{d}x$ associated with each track.  In addition to
the TPC, the Time-Of-Flight (TOF) detector was used for identification
of particles with transverse momentum~\pt~$>$~0.5~\gevc.

The TPC has 18 sectors covering full azimuth with 159 pad rows
radially placed in each sector. Tracks with at least 80 space points
in the TPC were used in this analysis.  Tracks compatible with a decay
in flight (kink topology) were rejected.  The track quality was
determined by the $\chi^{2}$ of the Kalman filter fit to the
reconstructed TPC clusters~\cite{Alme:2010ke}.  The $\chi^2$ per
degree of freedom was required to be less than 4.  For primary track
selection, only trajectories passing within 3.2~cm from the primary
vertex in the longitudinal direction and 2.4~cm in the transverse
direction were used.  Based on the specific ionization energy loss in
the TPC gas compared with the corresponding Bethe-Bloch curve, and the
time of flight in TOF, a probability for each track to be a pion,
kaon, proton, or electron was determined. Particles for which the pion
probability was the largest were used in this analysis. This resulted
in an overall purity above 95\%, with small contamination from
electrons in the region where the $\mathrm{d}E/\mathrm{d}x$ for the
two particle types overlap. Pions were selected in the pseudorapidity
range $|\eta|<0.8$ and 0.15 $<$ \pt$<$ 1.5~\gevc.

%=======================================================================
%\section{Correlation Function}

The correlation function $C({\bq})$ was calculated as
\begin{eqnarray}
  C({\bq})=\frac{ A({\bq})} { B({\bq})},
\end{eqnarray}
where $\bq=\bp_1-\bp_2$ is the relative momentum of two pions,
$A(\bq)$ is the distribution of particle pairs from the same event,
and $B(\bq)$ is the background distribution of uncorrelated particle
pairs. The background distribution is built by using the mixed-event
technique~\cite{Kopylov:1974th} in which pairs are made out of
particles from three different events with similar centrality (less
than 2\% difference), event-plane angle (less than 6$^\circ$
difference), and event vertex position along the beam direction (less
than 4~cm difference).  Both the $A({\bq})$ and $B({\bq})$
distributions were measured differentially with respect to the third
harmonic event-plane angle $\PsiTh$.  Note, that measurements relative
to $\PsiTh$ will smear any contribution from elliptic flow as the
elliptic and triangular event planes are
uncorrelated~\cite{ALICE:2011ab}.  The third harmonic event-plane
angle $\PsiTh$ was determined using TPC tracks.  To avoid
auto-correlation each event was split into two subevents
($-0.8<\eta<0$ and $0<\eta<0.8$). Pairs were chosen from one subevent
and the third harmonic event-plane angle $\PsiTh$ was estimated using
the particles from the other subevent, and vice-versa, with the event
plane resolution determined from the correlations between the event
planes determined in different
subevents~\cite{Poskanzer:1998yz}. Requiring a minimum value in the
two-track separation parameters $\Delta
\varphi^*=|\varphi^{*}_{1}-\varphi^{*}_{2}|$ and
$\Delta\eta=|\eta_{1}-\eta_{2}|$ reduces two-track reconstruction
effects such as track splitting or track merging.  The quantity
$\varphi^{*}$ is defined in this analysis as the azimuthal angle of
the track in the laboratory frame at the radial position of 1.6~m
inside the TPC. Splitting is the effect when one track is
reconstructed as two tracks, and merging is the effect of two tracks
being reconstructed as one. Also, to reduce the splitting effect,
pairs that share more than 5\% of the TPC clusters were removed from
the analysis. It is observed that at large relative momentum the
correlation function is a constant, and the background pair
distribution is normalized such that this constant is equal to unity.
The analysis was performed for different collision centralities in
several ranges of $k_\rT$, the magnitude of the pion-pair transverse
momentum $\bk_\rT=(\bp_{\rT,1}+\bp_{\rT,2})/2$, and in bins of $\dphi=
\mathrm{\varphi_{pair}}-\PsiTh$, where $\mathrm{\varphi_{pair}}$ is
the pair azimuthal angle. The
Bertsch-Pratt~\cite{Pratt:1986cc,Bertsch:1988db} out--side--long
coordinate system was used with the {\it long} direction pointing
along the beam axis, {\it out} along the transverse pair momentum, and
{\it side} being perpendicular to the other two.  The
three-dimensional correlation function was analyzed in the
Longitudinally Co-Moving System (LCMS)~\cite{LCM}, in which the total
longitudinal momentum of the pair is zero,
${p_{1,\mathrm{L}}}=-{p_{2,\mathrm{L}}}$.

To isolate the Bose--Einstein contribution in the correlation function,
effects due to final-state Coulomb repulsion must be taken into
account.  For that, the Bowler-Sinyukov fitting
procedure~\cite{Bowler:1991pi,Sinyukov:1998fc} was used in which the
Coulomb weight is only applied to the fraction of pairs ($\lambda$)
that participate in the Bose--Einstein correlation. In this approach,
the correlation function is fitted by
\begin{eqnarray}
 C({\bq},\dphi)=N[(1-\lambda)+\lambda K({\bq})(1+G({\bq},\dphi))],
\end{eqnarray}
where $N$ is the normalization factor. The function $G({\bq},\dphi)$
describes the Bose--Einstein correlations and $K({\bq})$ is the Coulomb
part of the two-pion wave function integrated over a source function
corresponding to $G({\bq})$.  In this analysis the Gaussian form of
$G({\bq},\dphi)$~\cite{guassain} was used
\begin{eqnarray}
 G(\bq,\dphi)=\exp
\left[
-\qout^{2} \rout^{2}(\dphi)-\qside^{2} \rside^{2}(\dphi) \right.
\nonumber
\\
\left.
-\qlong^{2} \rlong^{2}(\dphi)-2\qout \qside \ros^{2}(\dphi) \right.
\nonumber
\\
\left.
-2\qside \qlong \rsl^{2}(\dphi)-2\qout \qlong \rol^{2}(\dphi)
\right],
\end{eqnarray}
where the parameters $\rout$, $\rside$, and $\rlong$ are traditionally
called HBT radii in the {\it out}, {\it side}, and {\it long}
directions. The cross-terms $\ros^{2}$, $\rsl^{2}$, and $\rol^{2}$
describe the correlation in the {\it out}-{\it side}, {\it side}-{\it
  long}, and {\it out}-{\it long}~directions, respectively.

The systematic uncertainties on the extracted radii, discussed below,
vary in $\kt$ and centrality.  They include uncertainties related to
the tracking efficiency and track quality, momentum resolution,
different values for pair cuts ($\Delta \varphi^*$ and $\Delta\eta$),
and correlation function fit ranges~\cite{Adam:2015vna}.  Similarly to
the azimuthally inclusive analysis~\cite{Adam:2015vna}, different pair
cuts were used, with the default values chosen based on a Monte Carlo
study. The difference in the results from using different pair cuts
rather than the default pair cuts were included in the systematic
uncertainties (1--4\%). For different $\kt$ and centrality ranges,
different fitting ranges of correlation function were used as the
width of the correlation function depends on $\kt$ and centrality
range. The difference in the results from using different fit ranges
are due to the contamination of electrons in the particle
identification and the non-perfect Gaussian source (1--3\%). We also
studied the difference in the results by using positive and negative
pion pairs separately as well as data obtained with two opposite
magnetic field polarities of the ALICE L3 magnet. They have been
analyzed separately and a small difference in the results (less than
3\%) has been also accounted for in the systematic uncertainty. The
total systematic uncertainties were obtained by adding in quadrature
the contributions from all various sources mentioned above. The
systematic uncertainty associated with the event plane determination
is negligible compared to other sytematic uncertainties; the procedure
for the reaction plane resolution correction of the results is
described in the next section.

%=================================================================
\section{Results}

\begin{figure}[!ht]
\centering
\setlength\floatsep{0.25\baselineskip plus 3pt minus 2pt}
\setlength\textfloatsep{0.25\baselineskip plus 3pt minus 2pt}
\setlength\intextsep{0.25\baselineskip plus 3pt minus 2 pt}
\includegraphics[width=7cm,keepaspectratio]
{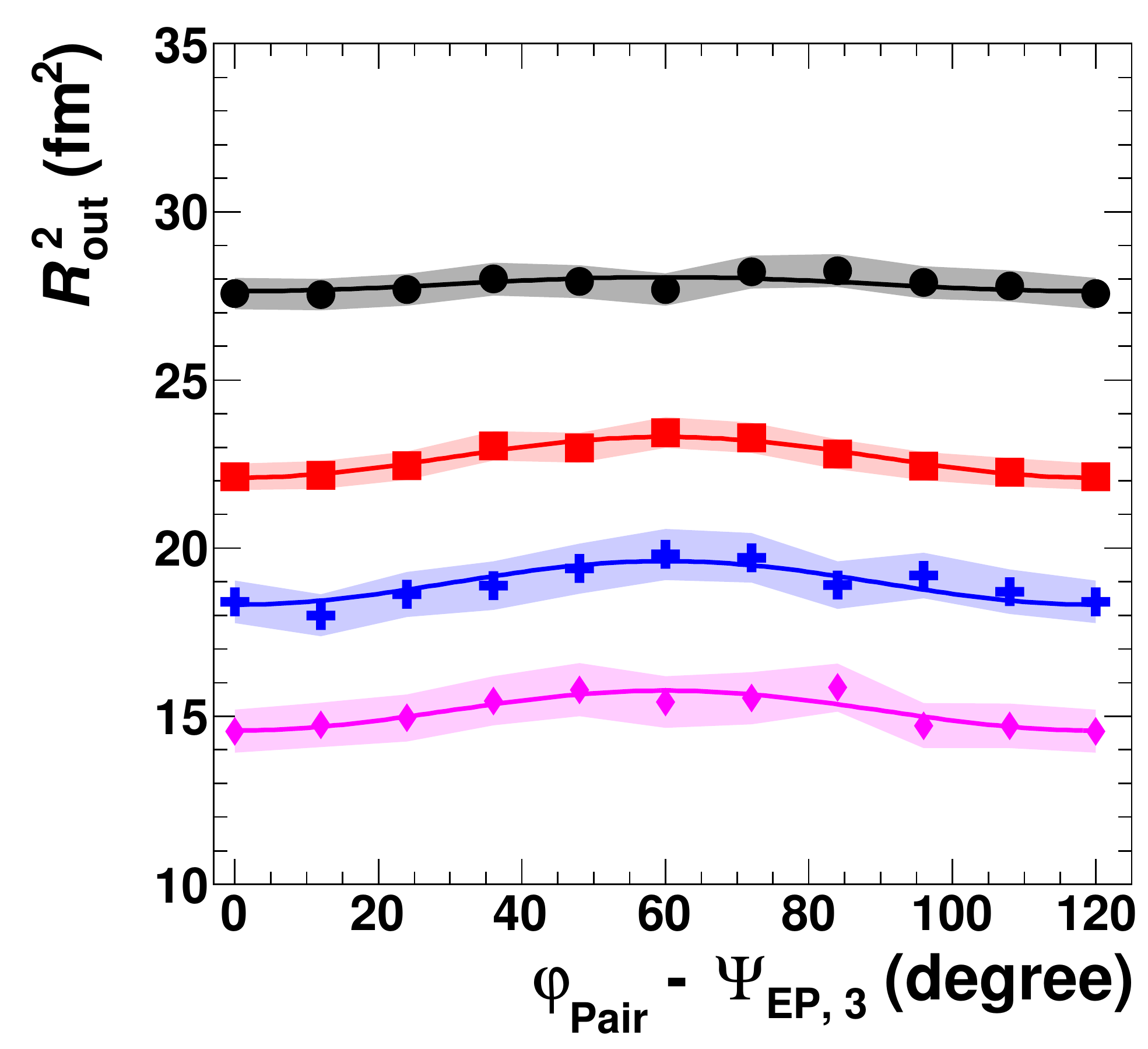}
\hspace{-0.2cm}
\includegraphics[width=7cm,keepaspectratio]{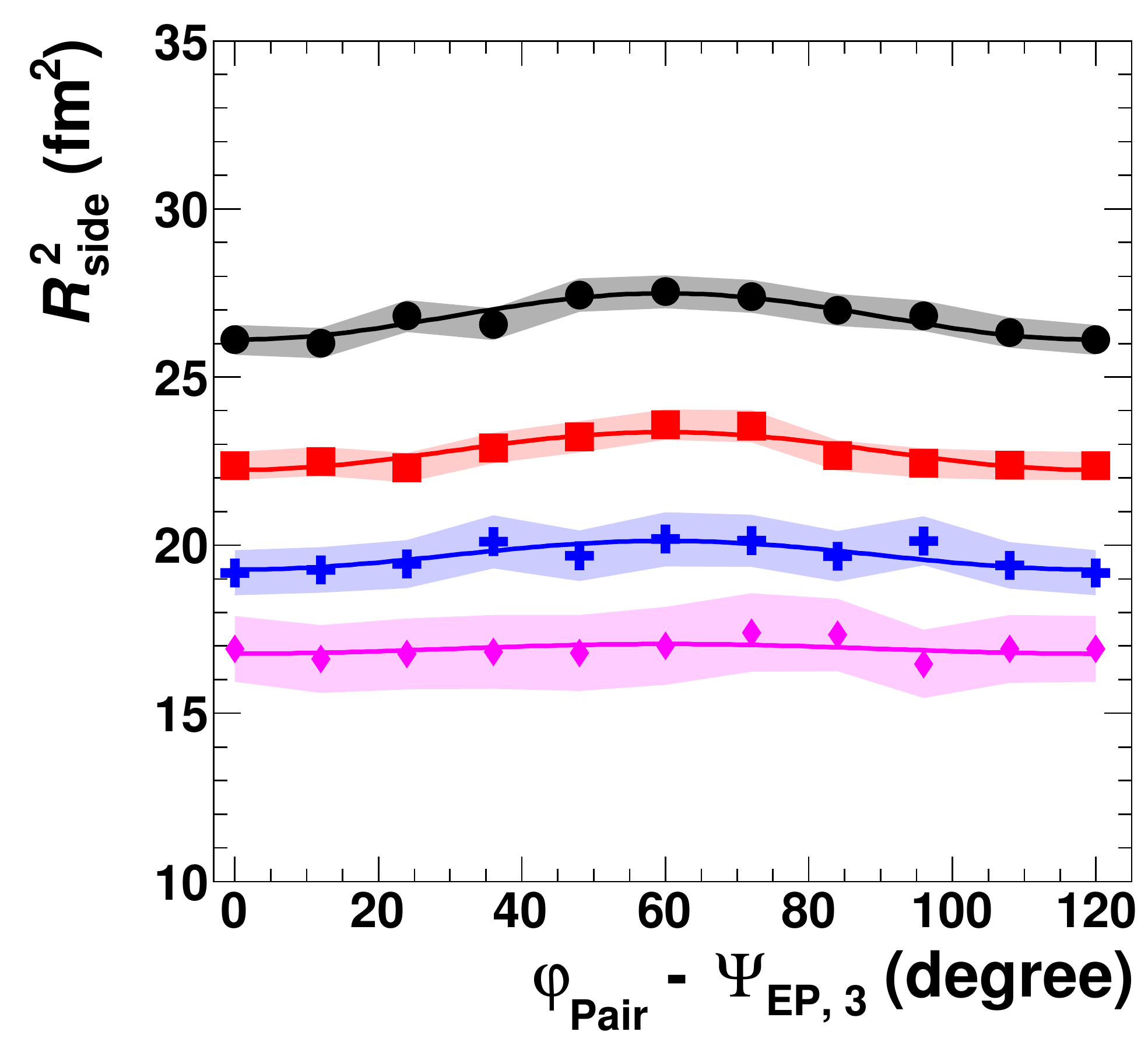}\\
\vspace{-0.6cm}
\includegraphics[width=7cm,keepaspectratio]{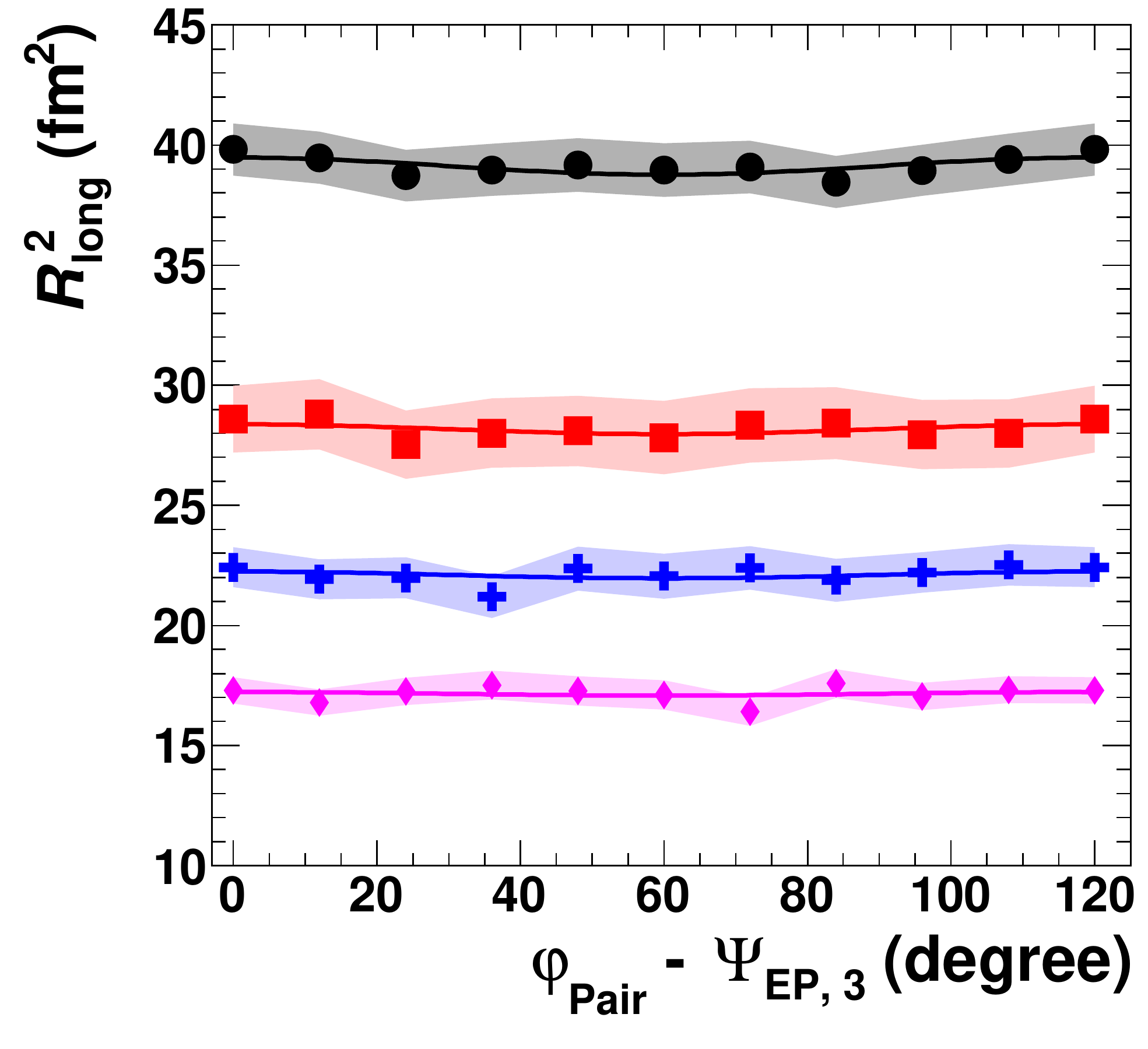}
\centering
\includegraphics[width=7cm,keepaspectratio]{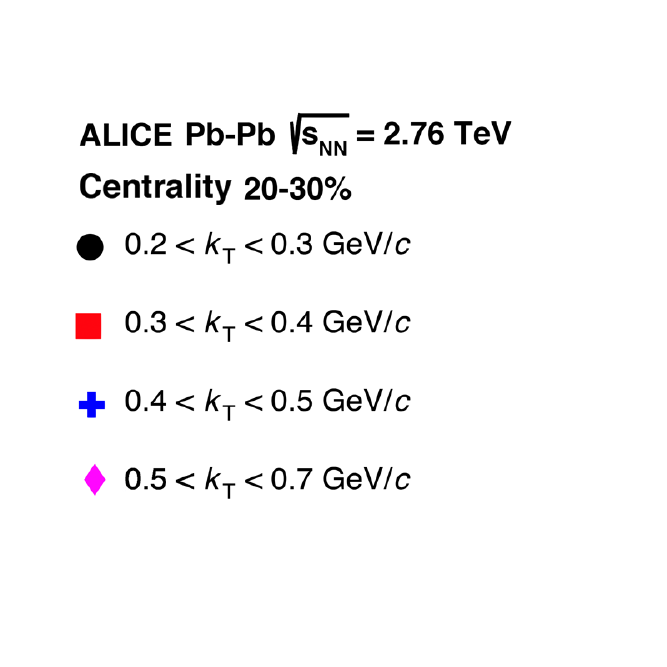}
  \caption{The azimuthal dependence of $R^{2}_{\rm out}$, $R^{2}_{\rm
      side}$, and $R^{2}_{\rm long}$ as a function of
    $\dphi=\mathrm{\varphi_{pair}}-\Psi_{3}$ for centrality percentile
    20--30\% and four different $\kt$ ranges. Solid lines represent
    the fit to the functional forms of Eq.~\ref{eq:radii_osc}. The
    shaded bands show the systematic uncertainty.  }
\label{fig:centdependence}
\end{figure}

Figure~\ref{fig:centdependence} presents the dependence of
$\rout^{2}$, $\rside^{2}$, and $\rlong^{2}$ on the pion emission angle
relative to the third harmonic event plane for centrality 20--30\% and
different $\kt$ ranges. Note that $\rout^2$ and $\rside^2$ exhibit
in-phase oscillations (for a quantitative analysis, see below).
Within the uncertainties of the measurement, $\rlong^2$ oscillations, if
any, are insignificant. Oscillations of $\rol^2$ and $\rsl^2$ radii
(not shown) are found to be consistent with zero, as expected due to
the source symmetry in longitudinal direction, and are not further
investigated. The curves represent the fits to the data using the
functions~\cite{Voloshin:2011mg}:
 \begin{eqnarray} \label{eq:radii_osc}
   \begin{split}
 R^{2}_{\mu}(\dphi)&=R^{2}_{\mu,0}+2R^{2}_{\mu,3}\cos(3\dphi)~~(\mu={\rm out,side,long}),
\\
\ros^{2}(\dphi)&=R^{2}_{{\rm os},0}+2 R^{2}_{{\rm os},3}\sin(3\dphi).
\end{split}
\end{eqnarray}

Fitting the radii's azimuthal dependence with the functional forms of
Eq.~\ref{eq:radii_osc}~allows us to extract the average radii and the
amplitudes of oscillations. The $\chi^{2}$ per number of degree of
freedom is 0.3--1.8 depending on $\kt$ and centrality range.  The
results for the average radii $R^{2}_{{\rm out},0}$, $R^{2}_{{\rm
    side},0}$, and $R^{2}_{{\rm os},0}$ were found to be consistent
with those reported previously in~\cite{Adamova:2017opl} in
azimuthally inclusive analysis.  The extracted amplitudes of
oscillations have to be corrected for the finite event plane
resolution. There exist several methods for such a
correction~\cite{Lisa:2005dd}, which produce consistent
results~\cite{Adamczyk:2014mxp} well within uncertainties of this
analysis. The results shown below have been obtained with the simplest
method first used by the E895 Collaboration~\cite{Lisa:2000xj}, in
which the amplitude of oscillation is divided by the event plane
resolution.  In this analysis the event plane resolution correction
factor is about 0.6--0.7, depending on centrality.
\begin{figure}[!ht]
\centering
\includegraphics[width=16cm,keepaspectratio]
{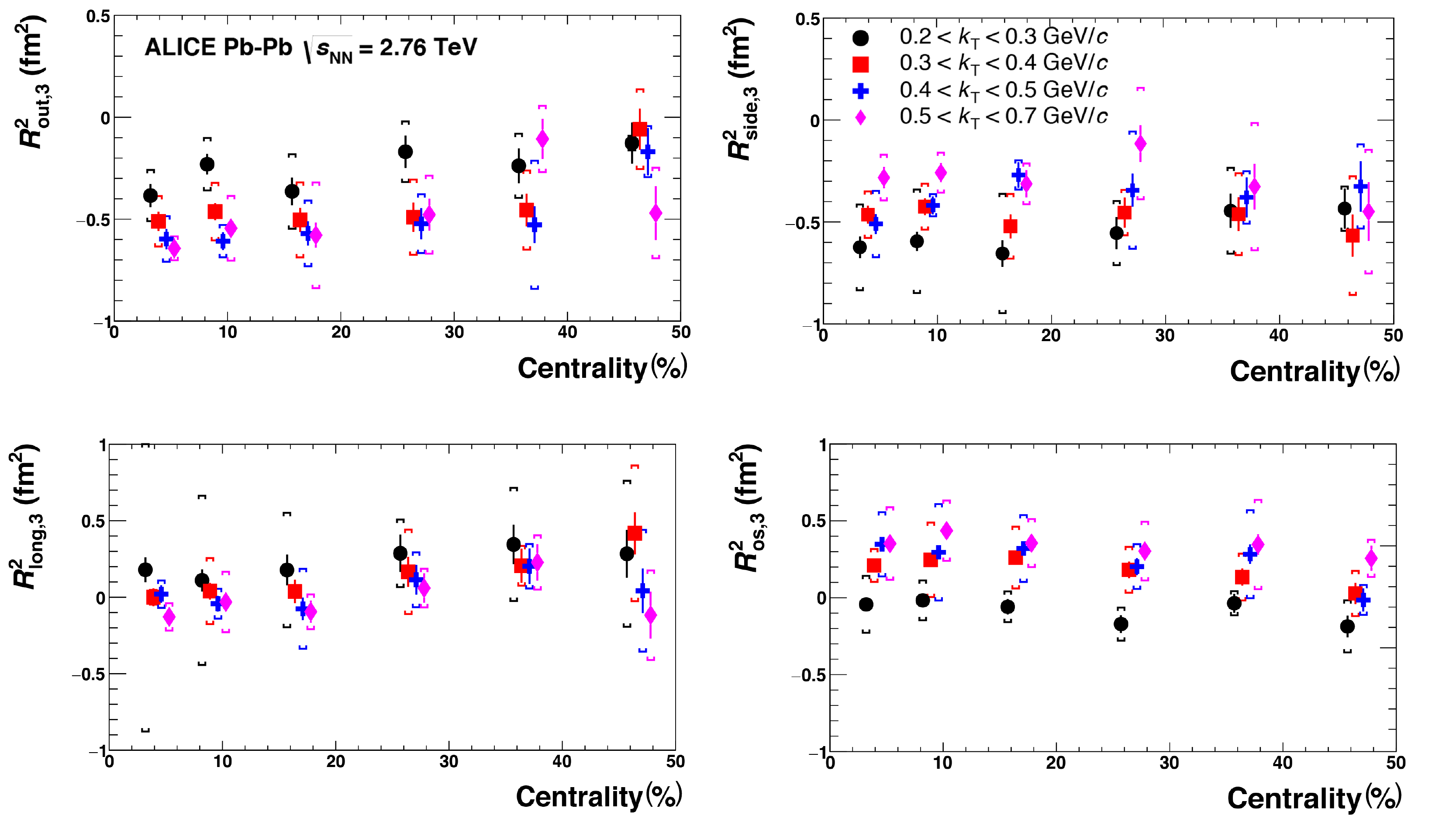}
  \caption{The amplitudes of radii oscillations~$R^{2}_{\rm
      out,3}$,~$R^{2}_{\rm side,3}$, $R^{2}_{\rm long,3}$,
    and~$R^{2}_{\rm os,3}$ versus centrality percentiles for four
    $\kt$ ranges. Square brackets indicate systematic uncertainties.
  }
\label{fig:Fig1}
\end{figure}

Figure \ref{fig:Fig1} shows the oscillation parameters $R^{2}_{{\rm
    out},3}$, $R^{2}_{{\rm side},3}$, $R^{2}_{{\rm long},3}$, and
$R^{2}_{{\rm os},3}$ for different centrality and $\kt$ ranges.  All
radii oscillations exhibit weak centrality dependence, likely
reflecting the weak centrality dependence of the triangular flow
itself.  The $\kt$ dependence is different for different radii
oscillations: while the magnitudes of $R^{2}_{{\rm out},3}$ and
$R^{2}_{{\rm os},3}$ are smallest for the smallest $\kt$ range, it is
opposite for $R^{2}_{{\rm side},3}$ (and, possibly for $R^{2}_{{\rm
    long},3}$), where the oscillations become stronger.  The parameter
$R^{2}_{{\rm long},3}$ is consistent with zero within the systematic
uncertainties while $R^{2}_{{\rm os},3}$ is positive for all
centralities and $\kt$ ranges except for the lowest $\kt$ range
0.2--0.3~\gevc.  Note that $R^{2}_{{\rm out},3}$ and $R^{2}_{{\rm
    side},3}$ are negative for all centralities and $\kt$ ranges.  In
the toy model simulations~\cite{Plumberg:2013nga} such phases of radii
oscillations were observed only in the so-called ``flow anisotropy
dominated case'' (a circular source with the radial expansion velocity
including the third harmonic modulation) and not for ``geometry
dominated'' case (triangular shape source with radial expansion
velocity proportional to radial distance from the center, with corners
having largest expansion velocity).
\begin{figure}[!ht]
\centering
\includegraphics[width=16cm,keepaspectratio]
{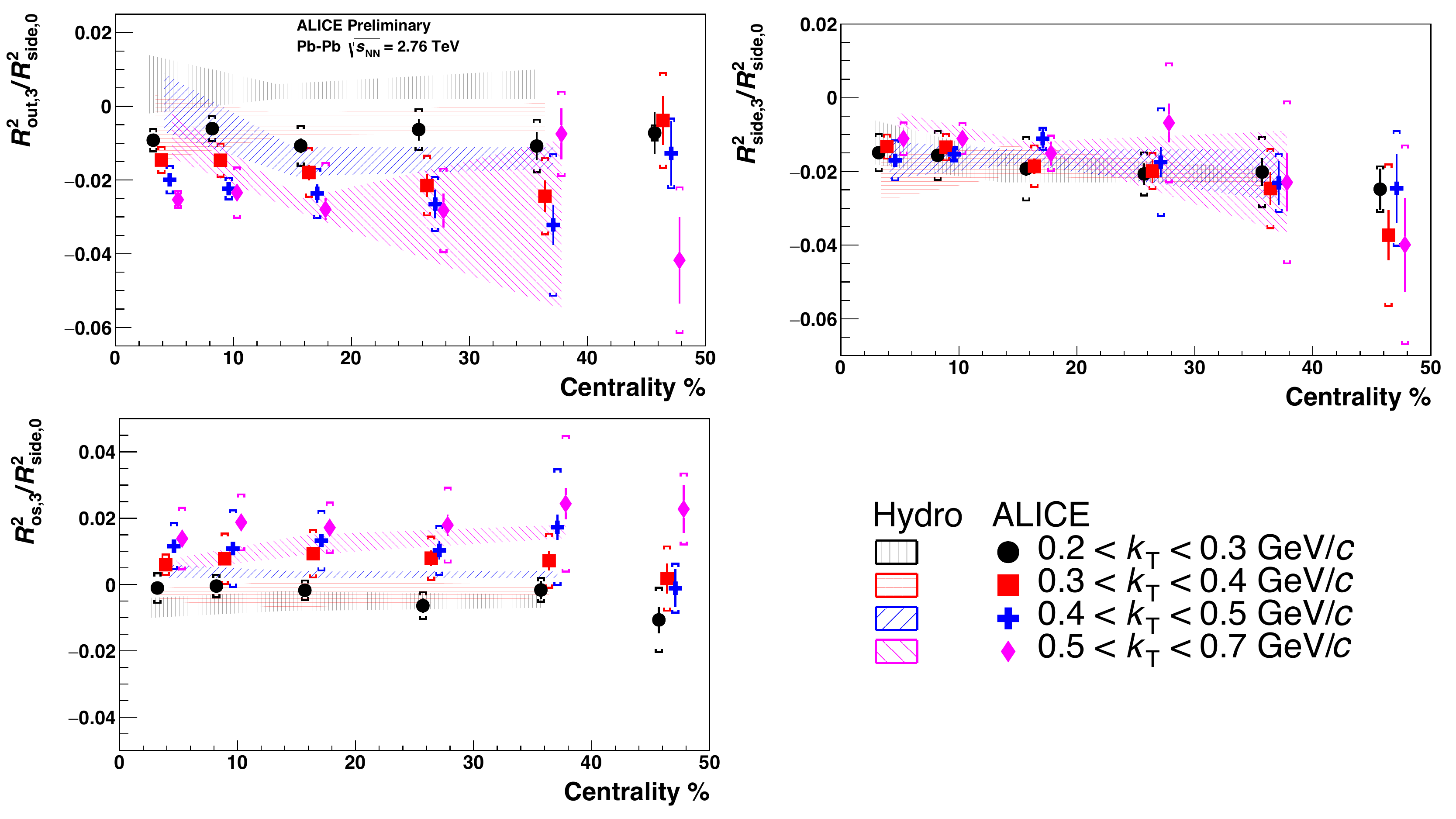}
  \caption{Amplitudes of the relative radii oscillations~$R^{2}_{\rm
      out,3}/R^{2}_{\rm side,0}$,~$R^{2}_{\rm side,3}/R^{2}_{\rm
      side,0}$, and~$R^{2}_{\rm os,2}/R^{2}_{\rm side,0}$ versus
    centrality for four $\kt$ ranges. Square brackets indicate
    systematic uncertainties. The shaded bands are the 3+1D
    hydrodynamical calculations~\cite{Bozek:2014hwa} and the width of
    the bands represent the uncertainties in the model
    calculations.
  }
\label{fig:Third}
\end{figure}

Figure \ref{fig:Third} shows the relative amplitudes of radius
oscillations $R^{2}_{{\rm out},3}/R^{2}_{{\rm side},0}$,~$R^{2}_{{\rm
    side},3}/R^{2}_{{\rm side},0}$, and $R^{2}_{{\rm
    os},3}/R^{2}_{{\rm side},0}$.  Similar to the previous analyses
and theoretical calculations~\cite{Bozek:2014hwa} we report all the
radii oscillations relative to the side radius the least affected by
the emission time duration. There exist no obvious centrality
dependence. As the average radii decrease with increasing $\kt$, the
$\kt$ dependence of relative oscillation amplitudes appear much
stronger for ``out'' and ``out-side'' radii, while ``side'' radius
relative amplitude exhibits no $\kt$ dependence with the
uncertainties.  The shaded bands in Fig.\ref{fig:Third} indicate the
results of 3+1D hydrodynamical calculations~\cite{Bozek:2014hwa}.
These calculations assume constant shear viscosity to entropy ratio
$\eta/s = 0.08$ and bulk viscosity that is nonzero in the hadronic
phase $\zeta /s = 0.04$, and the initial density from a Glauber Monte
Carlo model. The parameters of the model, were tuned to reproduce the
measured charged particle spectra, the elliptic and triangular flow.
We find that the relative amplitudes $R^{2}_{{\rm side},3}/R^{2}_{{\rm
    side},0}$ agree with these results rather well, while the relative
amplitudes $R^{2}_{{\rm out},3}/R^{2}_{{\rm side},0}$ and $R^{2}_{{\rm
    os},2}/R^{2}_{{\rm side},0}$ agree only qualitatively.  According
to the 3+1D hydrodynamical calculations, the negative signs of
$R^{2}_{{\rm side},3}$ and $R^{2}_{{\rm out},3}$ parameters are an
indication that the initial triangularity has been washed-out or even
reversed at freeze-out due to triangular flow~\cite{Bozek:2014hwa}.
\begin{figure}[!ht]
\centering
\includegraphics[width=10cm,keepaspectratio]
{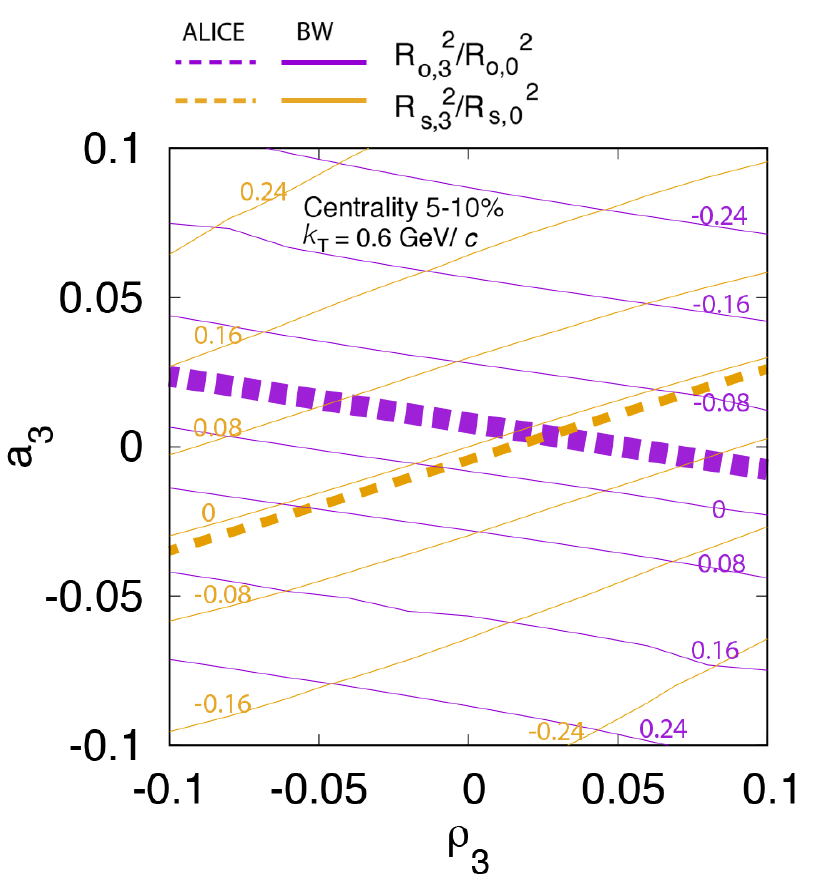}
  \caption{The relative amplitudes of the radius oscillations
    $R^{2}_{{\rm out},3}/R^{2}_{{\rm out},0}$, and~$R^{2}_{{\rm
        side},3}/R^{2}_{{\rm side},0}$ on the third-order anisotropies
    in space ($a_{3}$) and transverse flow ($\rho_{3}$) for the
    centrality range 5--10\% and $\kt$ = 0.6 \gevc~from the Blast-Wave
    model~\cite{Cimerman:2017lmm}. The thin dashed lines show the
    lines of a constant relative amplitude, in magenta for
    $R^{2}_{{\rm out},3}/R^{2}_{{\rm out},0}$ and in dark yellow for
    ~$R^{2}_{{\rm side},3}/R^{2}_{{\rm side},0}$. The thick lines show
    the corresponding ALICE results, with width of the lines
    representing the experimental uncertainties.  }
\label{fig:BW}
\end{figure}
\begin{figure}[!ht]
\centering
\includegraphics[width=10cm,keepaspectratio]
{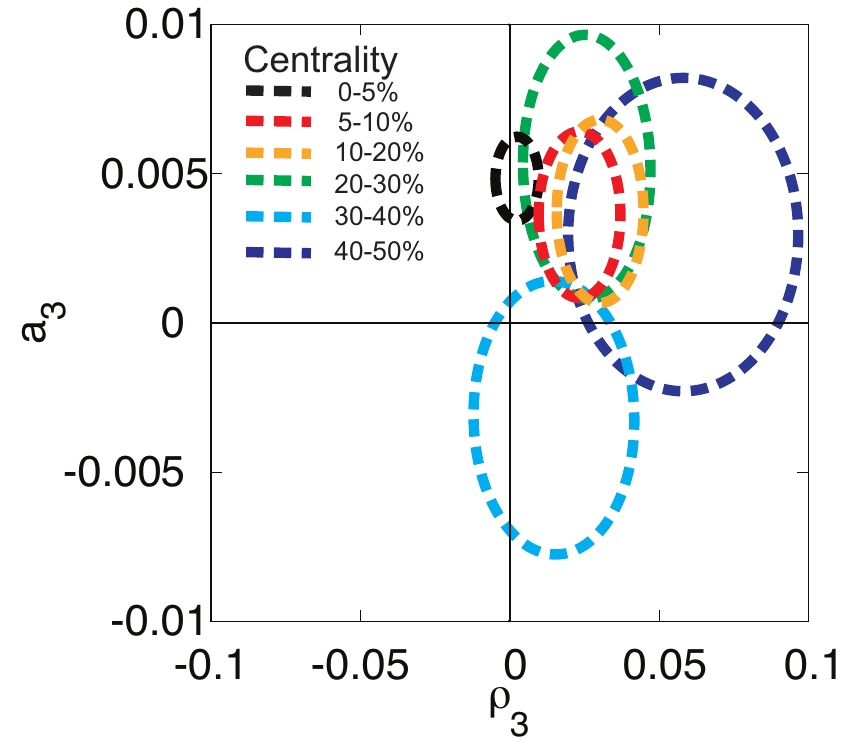}
  \caption{ Blast-Wave model~\cite{Cimerman:2017lmm} source
    parameters, final spatial ($a_{3}$) and transverse flow
    ($\rho_{3}$) anisotropies, for different centrality ranges, as
    obtained from the fit to ALICE radii oscillation parameters. The
    contours represent the one sigma uncertainty.  }
\label{fig:BW2}

\end{figure}

To investigate further on the final source shape, we compare our
results to the Blast-Wave model
calculations~\cite{Cimerman:2017lmm}. In that model, the spatial
geometry of the pion source at freeze-out is parameterized by
\begin{equation}
\label{e:Rtheta}
R(\phi) = R_0 \left ( 1 - \sum_{n=2}^\infty a_n \cos(n(\phi - \Psi_n))
\right )\, ,
\end{equation}
where $\Psi_{n}$'s denote the orientations of the $n$-th order symmetry
planes. The amplitudes $a_n$ and the phases $\Psi_n$ are model
parameters. The magnitude of the transverse expansion velocity is
parameterized as   $v_t=\tanh\rho$, where
the transverse rapidity
 $\rho$~\cite{Cimerman:2017lmm,Plumberg:2013nga} is 
\begin{equation}
\rho(\tilde{r},\phi_b) = \rho_{0} \, \tilde{r} \left (
1 + \sum_{n=2}^{\infty} 2 \rho_{n} \cos \left ( n (\phi_{b} - \Psi_{n})\right )
\right ).
\end{equation}
Here $\tilde{r}=r/R(\phi)$, and $\phi_b(\phi)$ is the transverse boost
direction assumed to be perpendicular to the surface of constant
$\tilde{r}$.  The results of this model presented below were obtained
assuming a kinetic freeze-out temperature of 120 MeV, and maximum
expansion rapidity $\rho_{0}$ = 0.8, tuned to describe single particle
spectra.  Figure~\ref{fig:BW} shows the relative amplitudes of the
radius oscillations $R^{2}_{{\rm out},3}/R^{2}_{{\rm out},0}$,
and~$R^{2}_{{\rm side},3}/R^{2}_{{\rm side},0}$ as a function of
Blast-wave model third-order parameters, spatial anisotropy $a_{3}$
and transverse flow anisotropy $\rho_{3}$. Thin dashed lines represent
the lines of constant relative amplitudes, with numbers next to lines
indicating the relative amplitude values.  Thick dashed lines show the
ALICE results for $R^{2}_{{\rm out},3}/R^{2}_{{\rm out},0}$ and
$R^{2}_{{\rm side},3}/R^{2}_{{\rm side},0}$ with the thickness of the
lines indicating the uncertainties. The intersection of the two dashed
lines corresponds to $a_{3}$ and $\rho_{3}$ parameters consistent with
ALICE measurements. The ALICE data and the Blast-Wave model
calculations correspond to pairs with $\kt$ = 0.6~\gevc~and the
centrality range 5--10\%. The comparison have been also performed for
other centralities and the corresponding Blast-Wave model parameters
have been deduced.  Figure~\ref{fig:BW2} presents the final source
spatial and transverse flow anisotropies for different centrality ranges from
matching the ALICE data with the Blast-Wave model calculations. The
contours correspond to one sigma uncertainty as derived from the fit
of the model to the data. It is observed that the final source
anisotropy is close to zero, significantly smaller than the initial
triangular eccentricities that are typically of the order of 0.2--0.3.
The negative values of the final source anisotropy would be
interpreted as that the triangular orientation at the initial-state is
reversed at freeze out.

\section{Summary}
We have reported a measurement of two-pion azimuthally-differential
femtoscopy relative to the third harmonic event plane in Pb--Pb
collisions at \snn~=~2.76 TeV. The observed osciallations of the HBT
radii unambiguously indicate a collective expansion of the system and
anisotropy in collective velocity fields at freeze-out.  Clear
in-phase oscillations of $R_{\rm out}$ and $R_{\rm side}$, with both
$R_{\rm out, 3}^2$ and $R^2_{\rm side, 3}$ parameters (as defined in
Eq.~\ref{eq:radii_osc}) being negative, have been observed for all
centralities and $\kt$ ranges.  According to model
calculations~\cite{Plumberg:2013nga} the observed $R_{\rm out}$ and
$R_{\rm side}$ in-phase oscillations are characteristics of the source
with strong triangular flow and close to zero spatial anisotropy. This
conclusion is further confirmed by a detailed comparison of our
results with the Blast-Wave model
calculations~\cite{Cimerman:2017lmm}, from which the parameters of the
source, the spatial anisotropy and modulations in the radial expansion
velocity, have been derived, with spatial triangular anisotropy being
more than an order of magnitude smaller than the typical initial
anisotropy values.  The oscillation amplitudes exhibit weak centrality
dependence, and in general decrease with decreasing $\kt$ except for
$R_{\rm side, 3}^2$ which on opposite is the largest in the smallest
$\kt$ bin.
The results of the 3+1D hydrodynamic
calculations~\cite{Bozek:2014hwa} are in a good qualitative agreement
with our measurements but, quantitatively, the model predicts a
stronger dependence of $R_{\rm out, 3}^2$ oscillations on $\kt$ than
observed in the data.

%
%
%

%%%%% acknowledgements
\newenvironment{acknowledgement}{\relax}{\relax}
\begin{acknowledgement}

\section*{Acknowledgements}

We thank J.~Cimerman and B.~Tomasik for providing us with the results of
the Blast-Model calculations~\cite{Cimerman:2017lmm}.
% Version: 2018-02-05

The ALICE Collaboration would like to thank all its engineers and technicians for their invaluable contributions to the construction of the experiment and the CERN accelerator teams for the outstanding performance of the LHC complex.
The ALICE Collaboration gratefully acknowledges the resources and support provided by all Grid centres and the Worldwide LHC Computing Grid (WLCG) collaboration.
The ALICE Collaboration acknowledges the following funding agencies for their support in building and running the ALICE detector:
A. I. Alikhanyan National Science Laboratory (Yerevan Physics Institute) Foundation (ANSL), State Committee of Science and World Federation of Scientists (WFS), Armenia;
Austrian Academy of Sciences and Nationalstiftung f\"{u}r Forschung, Technologie und Entwicklung, Austria;
Ministry of Communications and High Technologies, National Nuclear Research Center, Azerbaijan;
Conselho Nacional de Desenvolvimento Cient\'{\i}fico e Tecnol\'{o}gico (CNPq), Universidade Federal do Rio Grande do Sul (UFRGS), Financiadora de Estudos e Projetos (Finep) and Funda\c{c}\~{a}o de Amparo \`{a} Pesquisa do Estado de S\~{a}o Paulo (FAPESP), Brazil;
Ministry of Science \& Technology of China (MSTC), National Natural Science Foundation of China (NSFC) and Ministry of Education of China (MOEC) , China;
Ministry of Science and Education, Croatia;
Ministry of Education, Youth and Sports of the Czech Republic, Czech Republic;
The Danish Council for Independent Research | Natural Sciences, the Carlsberg Foundation and Danish National Research Foundation (DNRF), Denmark;
Helsinki Institute of Physics (HIP), Finland;
Commissariat \`{a} l'Energie Atomique (CEA) and Institut National de Physique Nucl\'{e}aire et de Physique des Particules (IN2P3) and Centre National de la Recherche Scientifique (CNRS), France;
Bundesministerium f\"{u}r Bildung, Wissenschaft, Forschung und Technologie (BMBF) and GSI Helmholtzzentrum f\"{u}r Schwerionenforschung GmbH, Germany;
General Secretariat for Research and Technology, Ministry of Education, Research and Religions, Greece;
National Research, Development and Innovation Office, Hungary;
Department of Atomic Energy Government of India (DAE), Department of Science and Technology, Government of India (DST), University Grants Commission, Government of India (UGC) and Council of Scientific and Industrial Research (CSIR), India;
Indonesian Institute of Science, Indonesia;
Centro Fermi - Museo Storico della Fisica e Centro Studi e Ricerche Enrico Fermi and Istituto Nazionale di Fisica Nucleare (INFN), Italy;
Institute for Innovative Science and Technology , Nagasaki Institute of Applied Science (IIST), Japan Society for the Promotion of Science (JSPS) KAKENHI and Japanese Ministry of Education, Culture, Sports, Science and Technology (MEXT), Japan;
Consejo Nacional de Ciencia (CONACYT) y Tecnolog\'{i}a, through Fondo de Cooperaci\'{o}n Internacional en Ciencia y Tecnolog\'{i}a (FONCICYT) and Direcci\'{o}n General de Asuntos del Personal Academico (DGAPA), Mexico;
Nederlandse Organisatie voor Wetenschappelijk Onderzoek (NWO), Netherlands;
The Research Council of Norway, Norway;
Commission on Science and Technology for Sustainable Development in the South (COMSATS), Pakistan;
Pontificia Universidad Cat\'{o}lica del Per\'{u}, Peru;
Ministry of Science and Higher Education and National Science Centre, Poland;
Korea Institute of Science and Technology Information and National Research Foundation of Korea (NRF), Republic of Korea;
Ministry of Education and Scientific Research, Institute of Atomic Physics and Romanian National Agency for Science, Technology and Innovation, Romania;
Joint Institute for Nuclear Research (JINR), Ministry of Education and Science of the Russian Federation and National Research Centre Kurchatov Institute, Russia;
Ministry of Education, Science, Research and Sport of the Slovak Republic, Slovakia;
National Research Foundation of South Africa, South Africa;
Centro de Aplicaciones Tecnol\'{o}gicas y Desarrollo Nuclear (CEADEN), Cubaenerg\'{\i}a, Cuba and Centro de Investigaciones Energ\'{e}ticas, Medioambientales y Tecnol\'{o}gicas (CIEMAT), Spain;
Swedish Research Council (VR) and Knut \& Alice Wallenberg Foundation (KAW), Sweden;
European Organization for Nuclear Research, Switzerland;
National Science and Technology Development Agency (NSDTA), Suranaree University of Technology (SUT) and Office of the Higher Education Commission under NRU project of Thailand, Thailand;
Turkish Atomic Energy Agency (TAEK), Turkey;
National Academy of  Sciences of Ukraine, Ukraine;
Science and Technology Facilities Council (STFC), United Kingdom;
National Science Foundation of the United States of America (NSF) and United States Department of Energy, Office of Nuclear Physics (DOE NP), United States of America.

\end{acknowledgement}

\bibliographystyle{utphys}
\bibliography{reference}

%%%%%%%% Bibliography (In case of using bibtex generate the bbl requested by arXiv)
%\bibliographystyle{utphys}   % Remember we use title in the biblio
%\bibliography{biblio}
%\input {bibliography.tex}

%%%%%%%%% appendix with author list
\newpage
\appendix

\section{The ALICE Collaboration}
\label{app:collab}
% Collaboration: CERN-LHC-ALICE
% Generation Date is 2018-Feb-05

% How to use:
%%%%%%%%% appendix with author list
%\appendix
%\section{The ALICE Collaboration}
%\label{app:collab}
%\input{Alice_Authorslist_XXXX-Axx-XX.tex}
\begingroup
\small
\begin{flushleft}
S.~Acharya\Irefn{org139}\And 
F.T.-.~Acosta\Irefn{org22}\And 
D.~Adamov\'{a}\Irefn{org94}\And 
J.~Adolfsson\Irefn{org81}\And 
M.M.~Aggarwal\Irefn{org98}\And 
G.~Aglieri Rinella\Irefn{org36}\And 
M.~Agnello\Irefn{org33}\And 
N.~Agrawal\Irefn{org48}\And 
Z.~Ahammed\Irefn{org139}\And 
S.U.~Ahn\Irefn{org77}\And 
S.~Aiola\Irefn{org144}
% ...
\And
\renewcommand\labelenumi{\textsuperscript{\theenumi}~}

\section*{Affiliation notes}
\renewcommand\theenumi{\roman{enumi}}
\begin{Authlist}
\item \Adef{org*}Deceased
\item \Adef{orgI}Dipartimento DET del Politecnico di Torino, Turin, Italy
\item \Adef{orgII}M.V. Lomonosov Moscow State University, D.V. Skobeltsyn Institute of Nuclear, Physics, Moscow, Russia
\item \Adef{orgIII}Department of Applied Physics, Aligarh Muslim University, Aligarh, India
\item \Adef{orgIV}Institute of Theoretical Physics, University of Wroclaw, Poland
\end{Authlist}

\section*{Collaboration Institutes}
\renewcommand\theenumi{\arabic{enumi}~}
\begin{Authlist}
\item \Idef{org1}A.I. Alikhanyan National Science Laboratory (Yerevan Physics Institute) Foundation, Yerevan, Armenia
\item \Idef{org2}Benem\'{e}rita Universidad Aut\'{o}noma de Puebla, Puebla, Mexico
\item \Idef{org3}Bogolyubov Institute for Theoretical Physics, National Academy of Sciences of Ukraine, Kiev, Ukraine
\item \Idef{org4}Bose Institute, Department of Physics  and Centre for Astroparticle Physics and Space Science (CAPSS), Kolkata, India
\item \Idef{org5}Budker Institute for Nuclear Physics, Novosibirsk, Russia
\item \Idef{org6}California Polytechnic State University, San Luis Obispo, California, United States
\item \Idef{org7}Central China Normal University, Wuhan, China
\item \Idef{org8}Centre de Calcul de l'IN2P3, Villeurbanne, Lyon, France
\item \Idef{org9}Centro de Aplicaciones Tecnol\'{o}gicas y Desarrollo Nuclear (CEADEN), Havana, Cuba
\item \Idef{org10}Centro de Investigaci\'{o}n y de Estudios Avanzados (CINVESTAV), Mexico City and M\'{e}rida, Mexico
\item \Idef{org11}Centro Fermi - Museo Storico della Fisica e Centro Studi e Ricerche ``Enrico Fermi', Rome, Italy
\item \Idef{org12}Chicago State University, Chicago, Illinois, United States
\item \Idef{org13}China Institute of Atomic Energy, Beijing, China
\item \Idef{org14}Chonbuk National University, Jeonju, Republic of Korea
\item \Idef{org15}Comenius University Bratislava, Faculty of Mathematics, Physics and Informatics, Bratislava, Slovakia
\item \Idef{org16}COMSATS Institute of Information Technology (CIIT), Islamabad, Pakistan
\item \Idef{org17}Creighton University, Omaha, Nebraska, United States
\item \Idef{org18}Department of Physics, Aligarh Muslim University, Aligarh, India
\item \Idef{org19}Department of Physics, Ohio State University, Columbus, Ohio, United States
\item \Idef{org20}Department of Physics, Pusan National University, Pusan, Republic of Korea
\item \Idef{org21}Department of Physics, Sejong University, Seoul, Republic of Korea
\item \Idef{org22}Department of Physics, University of California, Berkeley, California, United States
\item \Idef{org23}Department of Physics, University of Oslo, Oslo, Norway
\item \Idef{org24}Department of Physics and Technology, University of Bergen, Bergen, Norway
\item \Idef{org25}Dipartimento di Fisica dell'Universit\`{a} 'La Sapienza' and Sezione INFN, Rome, Italy
\item \Idef{org26}Dipartimento di Fisica dell'Universit\`{a} and Sezione INFN, Cagliari, Italy
\item \Idef{org27}Dipartimento di Fisica dell'Universit\`{a} and Sezione INFN, Trieste, Italy
\item \Idef{org28}Dipartimento di Fisica dell'Universit\`{a} and Sezione INFN, Turin, Italy
\item \Idef{org29}Dipartimento di Fisica e Astronomia dell'Universit\`{a} and Sezione INFN, Bologna, Italy
\item \Idef{org30}Dipartimento di Fisica e Astronomia dell'Universit\`{a} and Sezione INFN, Catania, Italy
\item \Idef{org31}Dipartimento di Fisica e Astronomia dell'Universit\`{a} and Sezione INFN, Padova, Italy
\item \Idef{org32}Dipartimento di Fisica `E.R.~Caianiello' dell'Universit\`{a} and Gruppo Collegato INFN, Salerno, Italy
\item \Idef{org33}Dipartimento DISAT del Politecnico and Sezione INFN, Turin, Italy
\item \Idef{org34}Dipartimento di Scienze e Innovazione Tecnologica dell'Universit\`{a} del Piemonte Orientale and INFN Sezione di Torino, Alessandria, Italy
\item \Idef{org35}Dipartimento Interateneo di Fisica `M.~Merlin' and Sezione INFN, Bari, Italy
\item \Idef{org36}European Organization for Nuclear Research (CERN), Geneva, Switzerland
\item \Idef{org37}Faculty of Engineering and Science, Western Norway University of Applied Sciences, Bergen, Norway
\item \Idef{org38}Faculty of Nuclear Sciences and Physical Engineering, Czech Technical University in Prague, Prague, Czech Republic
\item \Idef{org39}Faculty of Science, P.J.~\v{S}af\'{a}rik University, Ko\v{s}ice, Slovakia
\item \Idef{org40}Frankfurt Institute for Advanced Studies, Johann Wolfgang Goethe-Universit\"{a}t Frankfurt, Frankfurt, Germany
\item \Idef{org41}Gangneung-Wonju National University, Gangneung, Republic of Korea
\item \Idef{org42}Gauhati University, Department of Physics, Guwahati, India
\item \Idef{org43}Helmholtz-Institut f\"{u}r Strahlen- und Kernphysik, Rheinische Friedrich-Wilhelms-Universit\"{a}t Bonn, Bonn, Germany
\item \Idef{org44}Helsinki Institute of Physics (HIP), Helsinki, Finland
\item \Idef{org45}Hiroshima University, Hiroshima, Japan
\item \Idef{org46}Hochschule Worms, Zentrum  f\"{u}r Technologietransfer und Telekommunikation (ZTT), Worms, Germany
\item \Idef{org47}Horia Hulubei National Institute of Physics and Nuclear Engineering, Bucharest, Romania
\item \Idef{org48}Indian Institute of Technology Bombay (IIT), Mumbai, India
\item \Idef{org49}Indian Institute of Technology Indore, Indore, India
\item \Idef{org50}Indonesian Institute of Sciences, Jakarta, Indonesia
\item \Idef{org51}INFN, Laboratori Nazionali di Frascati, Frascati, Italy
\item \Idef{org52}INFN, Sezione di Bari, Bari, Italy
\item \Idef{org53}INFN, Sezione di Bologna, Bologna, Italy
\item \Idef{org54}INFN, Sezione di Cagliari, Cagliari, Italy
\item \Idef{org55}INFN, Sezione di Catania, Catania, Italy
\item \Idef{org56}INFN, Sezione di Padova, Padova, Italy
\item \Idef{org57}INFN, Sezione di Roma, Rome, Italy
\item \Idef{org58}INFN, Sezione di Torino, Turin, Italy
\item \Idef{org59}INFN, Sezione di Trieste, Trieste, Italy
\item \Idef{org60}Inha University, Incheon, Republic of Korea
\item \Idef{org61}Institut de Physique Nucl\'{e}aire d'Orsay (IPNO), Institut National de Physique Nucl\'{e}aire et de Physique des Particules (IN2P3/CNRS), Universit\'{e} de Paris-Sud, Universit\'{e} Paris-Saclay, Orsay, France
\item \Idef{org62}Institute for Nuclear Research, Academy of Sciences, Moscow, Russia
\item \Idef{org63}Institute for Subatomic Physics, Utrecht University/Nikhef, Utrecht, Netherlands
\item \Idef{org64}Institute for Theoretical and Experimental Physics, Moscow, Russia
\item \Idef{org65}Institute of Experimental Physics, Slovak Academy of Sciences, Ko\v{s}ice, Slovakia
\item \Idef{org66}Institute of Physics, Bhubaneswar, India
\item \Idef{org67}Institute of Physics of the Czech Academy of Sciences, Prague, Czech Republic
\item \Idef{org68}Institute of Space Science (ISS), Bucharest, Romania
\item \Idef{org69}Institut f\"{u}r Kernphysik, Johann Wolfgang Goethe-Universit\"{a}t Frankfurt, Frankfurt, Germany
\item \Idef{org70}Instituto de Ciencias Nucleares, Universidad Nacional Aut\'{o}noma de M\'{e}xico, Mexico City, Mexico
\item \Idef{org71}Instituto de F\'{i}sica, Universidade Federal do Rio Grande do Sul (UFRGS), Porto Alegre, Brazil
\item \Idef{org72}Instituto de F\'{\i}sica, Universidad Nacional Aut\'{o}noma de M\'{e}xico, Mexico City, Mexico
\item \Idef{org73}iThemba LABS, National Research Foundation, Somerset West, South Africa
\item \Idef{org74}Johann-Wolfgang-Goethe Universit\"{a}t Frankfurt Institut f\"{u}r Informatik, Fachbereich Informatik und Mathematik, Frankfurt, Germany
\item \Idef{org75}Joint Institute for Nuclear Research (JINR), Dubna, Russia
\item \Idef{org76}Konkuk University, Seoul, Republic of Korea
\item \Idef{org77}Korea Institute of Science and Technology Information, Daejeon, Republic of Korea
\item \Idef{org78}KTO Karatay University, Konya, Turkey
\item \Idef{org79}Laboratoire de Physique Subatomique et de Cosmologie, Universit\'{e} Grenoble-Alpes, CNRS-IN2P3, Grenoble, France
\item \Idef{org80}Lawrence Berkeley National Laboratory, Berkeley, California, United States
\item \Idef{org81}Lund University Department of Physics, Division of Particle Physics, Lund, Sweden
\item \Idef{org82}Nagasaki Institute of Applied Science, Nagasaki, Japan
\item \Idef{org83}Nara Women{'}s University (NWU), Nara, Japan
\item \Idef{org84}National and Kapodistrian University of Athens, School of Science, Department of Physics , Athens, Greece
\item \Idef{org85}National Centre for Nuclear Research, Warsaw, Poland
\item \Idef{org86}National Institute of Science Education and Research, HBNI, Jatni, India
\item \Idef{org87}National Nuclear Research Center, Baku, Azerbaijan
\item \Idef{org88}National Research Centre Kurchatov Institute, Moscow, Russia
\item \Idef{org89}Niels Bohr Institute, University of Copenhagen, Copenhagen, Denmark
\item \Idef{org90}Nikhef, National institute for subatomic physics, Amsterdam, Netherlands
\item \Idef{org91}NRC ¿Kurchatov Institute¿ ¿ IHEP , Protvino, Russia
\item \Idef{org92}NRNU Moscow Engineering Physics Institute, Moscow, Russia
\item \Idef{org93}Nuclear Physics Group, STFC Daresbury Laboratory, Daresbury, United Kingdom
\item \Idef{org94}Nuclear Physics Institute of the Czech Academy of Sciences, \v{R}e\v{z} u Prahy, Czech Republic
\item \Idef{org95}Oak Ridge National Laboratory, Oak Ridge, Tennessee, United States
\item \Idef{org96}Petersburg Nuclear Physics Institute, Gatchina, Russia
\item \Idef{org97}Physics department, Faculty of science, University of Zagreb, Zagreb, Croatia
\item \Idef{org98}Physics Department, Panjab University, Chandigarh, India
\item \Idef{org99}Physics Department, University of Jammu, Jammu, India
\item \Idef{org100}Physics Department, University of Rajasthan, Jaipur, India
\item \Idef{org101}Physikalisches Institut, Eberhard-Karls-Universit\"{a}t T\"{u}bingen, T\"{u}bingen, Germany
\item \Idef{org102}Physikalisches Institut, Ruprecht-Karls-Universit\"{a}t Heidelberg, Heidelberg, Germany
\item \Idef{org103}Physik Department, Technische Universit\"{a}t M\"{u}nchen, Munich, Germany
\item \Idef{org104}Research Division and ExtreMe Matter Institute EMMI, GSI Helmholtzzentrum f\"ur Schwerionenforschung GmbH, Darmstadt, Germany
\item \Idef{org105}Rudjer Bo\v{s}kovi\'{c} Institute, Zagreb, Croatia
\item \Idef{org106}Russian Federal Nuclear Center (VNIIEF), Sarov, Russia
\item \Idef{org107}Saha Institute of Nuclear Physics, Kolkata, India
\item \Idef{org108}School of Physics and Astronomy, University of Birmingham, Birmingham, United Kingdom
\item \Idef{org109}Secci\'{o}n F\'{\i}sica, Departamento de Ciencias, Pontificia Universidad Cat\'{o}lica del Per\'{u}, Lima, Peru
\item \Idef{org110}Shanghai Institute of Applied Physics, Shanghai, China
\item \Idef{org111}Stefan Meyer Institut f\"{u}r Subatomare Physik (SMI), Vienna, Austria
\item \Idef{org112}SUBATECH, IMT Atlantique, Universit\'{e} de Nantes, CNRS-IN2P3, Nantes, France
\item \Idef{org113}Suranaree University of Technology, Nakhon Ratchasima, Thailand
\item \Idef{org114}Technical University of Ko\v{s}ice, Ko\v{s}ice, Slovakia
\item \Idef{org115}Technische Universit\"{a}t M\"{u}nchen, Excellence Cluster 'Universe', Munich, Germany
\item \Idef{org116}The Henryk Niewodniczanski Institute of Nuclear Physics, Polish Academy of Sciences, Cracow, Poland
\item \Idef{org117}The University of Texas at Austin, Austin, Texas, United States
\item \Idef{org118}Universidad Aut\'{o}noma de Sinaloa, Culiac\'{a}n, Mexico
\item \Idef{org119}Universidade de S\~{a}o Paulo (USP), S\~{a}o Paulo, Brazil
\item \Idef{org120}Universidade Estadual de Campinas (UNICAMP), Campinas, Brazil
\item \Idef{org121}Universidade Federal do ABC, Santo Andre, Brazil
\item \Idef{org122}University College of Southeast Norway, Tonsberg, Norway
\item \Idef{org123}University of Cape Town, Cape Town, South Africa
\item \Idef{org124}University of Houston, Houston, Texas, United States
\item \Idef{org125}University of Jyv\"{a}skyl\"{a}, Jyv\"{a}skyl\"{a}, Finland
\item \Idef{org126}University of Liverpool, Liverpool, United Kingdom
\item \Idef{org127}University of Split, Faculty of Electrical Engineering, Mechanical Engineering and Naval Architecture, Split, Croatia
\item \Idef{org128}University of Tennessee, Knoxville, Tennessee, United States
\item \Idef{org129}University of the Witwatersrand, Johannesburg, South Africa
\item \Idef{org130}University of Tokyo, Tokyo, Japan
\item \Idef{org131}University of Tsukuba, Tsukuba, Japan
\item \Idef{org132}Universit\'{e} Clermont Auvergne, CNRS/IN2P3, LPC, Clermont-Ferrand, France
\item \Idef{org133}Universit\'{e} de Lyon, Universit\'{e} Lyon 1, CNRS/IN2P3, IPN-Lyon, Villeurbanne, Lyon, France
\item \Idef{org134}Universit\'{e} de Strasbourg, CNRS, IPHC UMR 7178, F-67000 Strasbourg, France, Strasbourg, France
\item \Idef{org135} Universit\'{e} Paris-Saclay Centre d¿\'Etudes de Saclay (CEA), IRFU, Department de Physique Nucl\'{e}aire (DPhN), Saclay, France
\item \Idef{org136}Universit\`{a} degli Studi di Pavia, Pavia, Italy
\item \Idef{org137}Universit\`{a} di Brescia, Brescia, Italy
\item \Idef{org138}V.~Fock Institute for Physics, St. Petersburg State University, St. Petersburg, Russia
\item \Idef{org139}Variable Energy Cyclotron Centre, Kolkata, India
\item \Idef{org140}Warsaw University of Technology, Warsaw, Poland
\item \Idef{org141}Wayne State University, Detroit, Michigan, United States
\item \Idef{org142}Westf\"{a}lische Wilhelms-Universit\"{a}t M\"{u}nster, Institut f\"{u}r Kernphysik, M\"{u}nster, Germany
\item \Idef{org143}Wigner Research Centre for Physics, Hungarian Academy of Sciences, Budapest, Hungary
\item \Idef{org144}Yale University, New Haven, Connecticut, United States
\item \Idef{org145}Yonsei University, Seoul, Republic of Korea
\end{Authlist}
\endgroup
  %%%%%%% done by webmaster team
\end{document}